\documentclass[aps,prl,twocolumn,groupedaddress,showpacs]{revtex4}
\usepackage{graphicx}
\graphicspath{{figure file fold path}}
\usepackage[dvipsnames,usenames]{color}
\usepackage{color}
\usepackage{amsmath}
\usepackage{float}
\usepackage{amssymb}
\usepackage{amsthm}
\usepackage{mathrsfs}

\emergencystretch=\maxdimen
\begin{document}
\title{Deterministic Control of Extreme Events in a semiconductor VCSEL via Polarization-Engineered Optical Feedback}
\author{Tao Wang$^{1}$, Zhibo Li$^{1}$, Yixing Ma$^{1}$, Juncheng Huang$^{1}$, Yiheng Li$^{1}$, Zhicong Tu$^{1}$, Shuiying Xiang$^{1}$, Giancarlo Ruocco$^{2,3}$, and Yue Hao$^{4}$}

\affiliation{$^1$State Key Laboratory of Integrated Service Networks, Xidian University, Xi'an 710071, China}
\affiliation{$^2$Department of Physics, Sapienza University, Piazzale Aldo Moro 5, 00185, Rome, Italy}
\affiliation{$^3$Center for Life Nano- $\&$ Neuro-Science, Italian Institute of Technology, Rome, Italy}
\affiliation{$^4$State Key Discipline Laboratory of Wide Bandgap Semiconductor Technology, School of Microelectronics, Xidian University, Xi’an 710071, China}

\date{\today}

\begin{abstract}
Extreme events, or rogue waves, are high-amplitude, rare occurrences that emerge across diverse physical systems and often defy conventional statistical predictions. While optical systems provide a controlled setting for studying these phenomena, achieving deterministic control over their generation remains challenging. Here, we demonstrate a novel approach to induce and precisely modulate extreme events in a semiconductor VCSEL using polarization-controlled optical feedback. By integrating a $\lambda$/2-waveplate into a polarization-selective external cavity, we regulate the nonlinear interaction between TE and TM modes. This setup triggers high-intensity, heavy-tailed fluctuations in the TM mode, exhibiting clear signatures of extreme events. We show that these events arise from deterministic energy exchange between modes, as evidenced by strong bipolar correlations and long-range temporal memory. The waveplate angle serves as an effective external parameter, enabling non-monotonic tuning of the event rate, intensity, and temporal clustering. Our study establish a platform for exploring extreme events in dissipative systems, with implications for nonlinear photonics and optical technologies.
\end{abstract}


\maketitle 

\section{Introduction and objectives}
Extreme events, also known as rogue or freak waves, are rare, isolated occurrences characterized by amplitudes that vastly exceed those of typical events~\cite{birkholz2015}. Their emergence defies predictions based on standard statistical distributions of background fluctuations. First identified in oceanography~\cite{chalikov2009freak, teague2007observed}, the concept of extreme events has proven universally relevant across diverse fields including finance, optics, and condensed matter physics~\cite{faranda2024statistical, xiao2024memory, wu2024control, hou2024predicting}. Their profound potential impact drives considerable interest in understanding and mitigating their effects~\cite{Cavalcante2013, spitz2020extreme, stroganov2024extreme, wang2024reservoir}.

In optics, the study of optical rogue waves provides a controllable and accessible testbed for investigating the universal physics underlying extreme events~\cite{Solli2007, dudley2019rogue}. These intense, transient pulses of light are manifestations of complexity in nonlinear systems, yet a complete understanding of their generation and dynamics remains an active pursuit. A critical enabler for these studies has been the development of real-time measurement techniques capable of capturing large datasets of these rare, ultrafast events~\cite{jalali2010real, walczak2017extreme}.

Dynamically, extreme events can be classified into several categories: (i) those arising from purely deterministic, classical dynamics in nonlinear systems~\cite{Nicolis2006, Bonatto2011}; (ii) those in classical systems driven by noise~\cite{Solli2007, hernandez2021noise}; and (iii) those predicted in quantum systems~\cite{carollo2018making}. While early research focused on integrable and conservative systems like nonlinear optical fibers~\cite{Solli2007, narhi2018machine}, recent interest has surged in observing these events within active, dissipative systems, particularly lasers~\cite{wang2024reservoir, ge2024enhanced}. In such non-equilibrium environments, a pivotal and unresolved question concerns the role of noise: is it merely a background perturbation, or a fundamental driver that enhances the probability of extreme events?

Semiconductor lasers, particularly Vertical-Cavity Surface-Emitting Lasers (VCSELs), are ideal systems for such studies due to their rich nonlinear dynamics~\cite{bittner2022complex} and sensitivity to optical feedback~\cite{akhmediev2016roadmap}. While optical feedback in edge-emitting lasers has been extensively studied for extreme events~\cite{ge2024characterizing, karsaklian2013extreme}, the unique polarization properties of VCSELs offer a distinct pathway for control. However, a significant challenge has been the lack of a simple, precise, and continuously tunable external parameter to deterministically influence the nonlinear interactions responsible for extreme event generation, which is significant for both fundamental understanding and potential applications.

In this work, we address this challenge by exploiting the polarization dynamics of a VCSEL subjected to engineered orthogonal polarization injection. We demonstrate that incorporating a half-wave ($\lambda$/2) plate within a polarization-selective feedback cavity provides an unprecedented level of control over the polarization state of the reinjected light. This allows us to finely tune the nonlinear competition between the transverse electric (TE) and transverse magnetic (TM) modes, successfully inducing and controlling extreme events in the TM channel. Our results provide a novel and versatile platform for extreme event studies and offer fundamental insights with implications for nonlinear laser physics and advanced photonic technologies.

\section{Experimental setup}
\begin{figure}[ht!]
\centering
  \includegraphics[width=8.5cm]{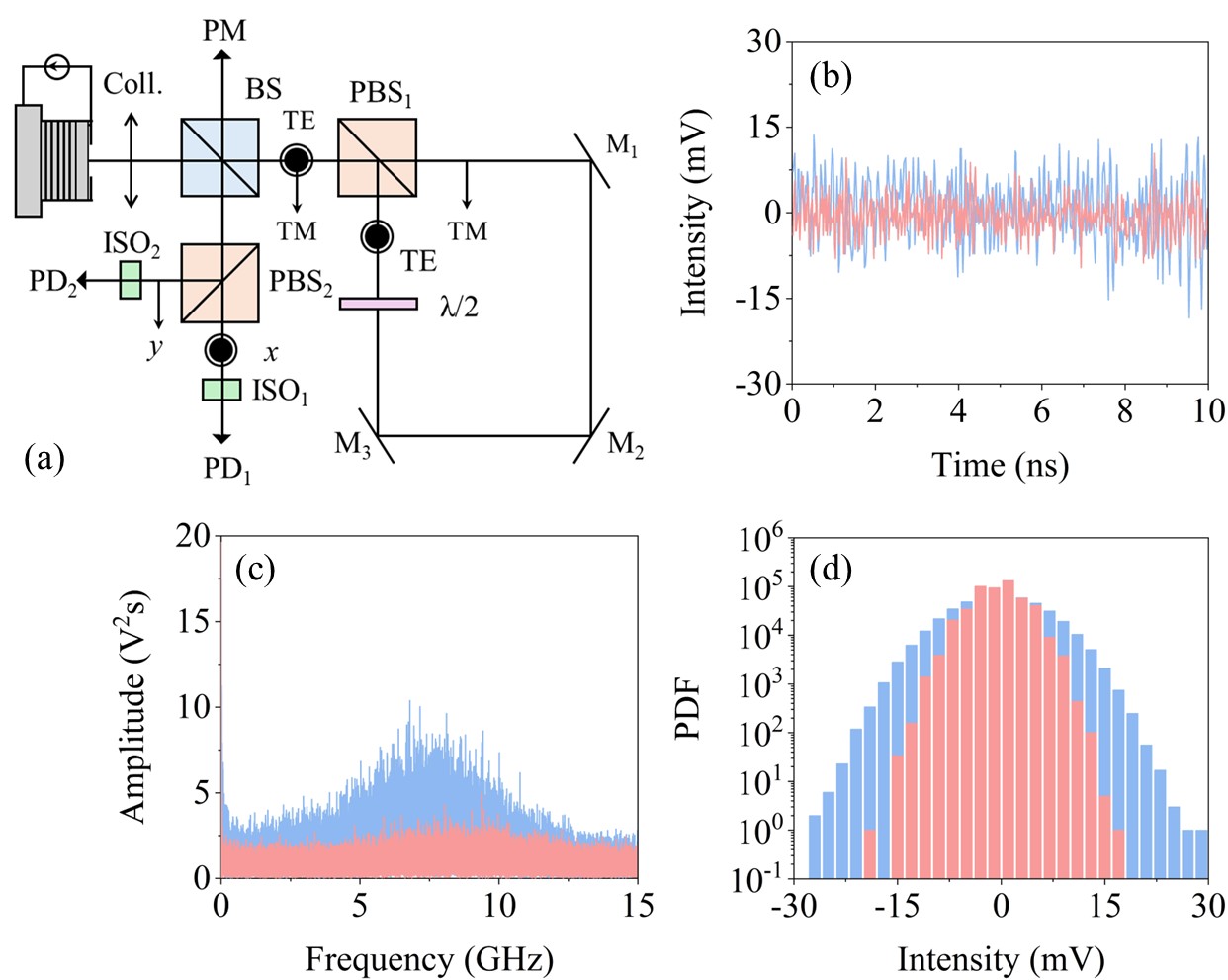}
  \caption{Experimental setup and fundamental characterizations: (a) Experimental setup. Coll., Collimator; BS, non-polarizing beamsplitter; PBS$_1$ and PBS$_2$, polarizing beamsplitters; M$_1$, M$_2$ and M$_3$, high-reflectivity mirrors; $\lambda$/2, half-wave plate; ISO$_1$ and ISO$_2$, optical isolators; PD$_1$ and PD$_2$, fast photodetectors; PM, power meter. (b) Temporal dynamics within 10 ns of the TE and TM modes of the VCSEL under the condition of free running and J = 3.00 mA. (c) the corresponding RF spectra of the TE and TM modes. (d) Intensity distribution histograms of the TE and TM modes.}
  \label{Setup_freerunning}
\end{figure}

The experimental setup, illustrated in Fig. \ref{Setup_freerunning}a, builds upon configurations used in previous studies~\cite{wang2025fast, wang2025mode}. Light emitted from a light source is first collimated and then directed through a non-polarizing 50:50 beam splitter (BS). This BS separates the beam into two paths: one for monitoring the laser output and the other for injection into the external feedback cavity.

The external cavity is constructed using a polarization beam splitter (PBS$_1$) and three high-reflectivity mirrors (M$_1$, M$_2$, and M$_3$). PBS$_1$ separates the orthogonal linear polarization states, routing the transverse electric (TE) and transverse magnetic (TM) modes into clockwise and counter-clockwise paths, respectively. The main element is a $\lambda$/2-waveplate, which precisely controls the intensity and polarization state of the light reinjected into the laser. After completing this loop, the light is fed back into the laser, providing a controlled perturbation.

The output TE and TM modes are subsequently separated by a second polarization beam splitter (PBS$_2$) and detected using two fast, AC-coupled photodiodes (PD$_1$ and PD$_2$, 10 GHz bandwidth). The signals are recorded by a digital oscilloscope (Tektronix DSA72004, 20 GHz bandwidth) operating at a sampling rate of 50 GSa/s. Optical isolators placed directly before the photodiodes are critical to maintain signal integrity by preventing back-reflections into the laser cavity. A power meter (PM) positioned after the initial BS monitors the feedback strength. 

The light source is a semiconductor VCSEL (Thorlabs L850VH1) operating at a wavelength of $\lambda = 850$ nm. The laser is driven by a low-noise current source (Thorlabs LDC205C), and its temperature is stabilized at $25^\circ\mathrm{C}$ using a dedicated controller (Thorlabs TED200C). The VCSEL operates in a single longitudinal and spatial mode, exhibiting a circular beam profile.

The total feedback path length is 1.53 m, corresponding to a roundtrip time of approximately 10.2 ns. As demonstrated in our previous study~\cite{wang2025fast}, the polarization conversion mechanism involves TE light transforming to TM polarization after passing through the waveplate, reflecting from the birefringent VCSEL cavity, and subsequently reconverting back to TE polarization before reinjection.

Fig.~\ref{Setup_freerunning}b displays the temporal dynamics of the TE and TM modes from the VCSEL under free-running conditions at a pump current of $J = 3.00$ mA. The TE mode (blue curve), which is the dominant lasing mode above threshold, exhibits intense, high-frequency oscillations characteristic of relaxation oscillations, with a characteristic frequency of approximately 7.5 GHz. This dynamical signature is corroborated by a pronounced peak in the corresponding radio frequency (RF) spectrum (blue curve, Fig.~\ref{Setup_freerunning}c). In contrast, the TM mode (red curve) operates below its lasing threshold at this pump current. Consequently, its intensity remains significantly weaker, fluctuating near zero with an smaller amplitude than that of the TE mode. The associated RF spectrum of the TM mode (red curve, Fig.~\ref{Setup_freerunning}c) lacks any distinct resonant peak and is instead dominated by a flat, broadband noise background, a hallmark of amplified spontaneous emission (ASE) in a sub-threshold gain regime~\cite{wang2025mode, henry2003theory}. 

The underlying statistics of these dynamics are further elucidated by the intensity distribution histograms presented in Fig.~\ref{Setup_freerunning}d. Both modes exhibit nearly symmetric distributions centred at zero, consistent with the AC-coupled detection of oscillatory signals. However, their profiles are markedly different. The TE mode (blue) displays a broad distribution, reflecting the large-amplitude intensity fluctuations driven by its inherent relaxation oscillation dynamics~\cite{coldren2012diode}. Conversely, the TM mode (red) exhibits a significantly narrower and more centralized distribution, which aligns with its non-lasing, sub-threshold state where intensity noise is substantially smaller and governed primarily by spontaneous emission processes.

\section{Results and Discussions}
\subsection{System Characterization under Optical Feedback}
\begin{figure}[ht!]
\centering
  \includegraphics[width=8.5cm]{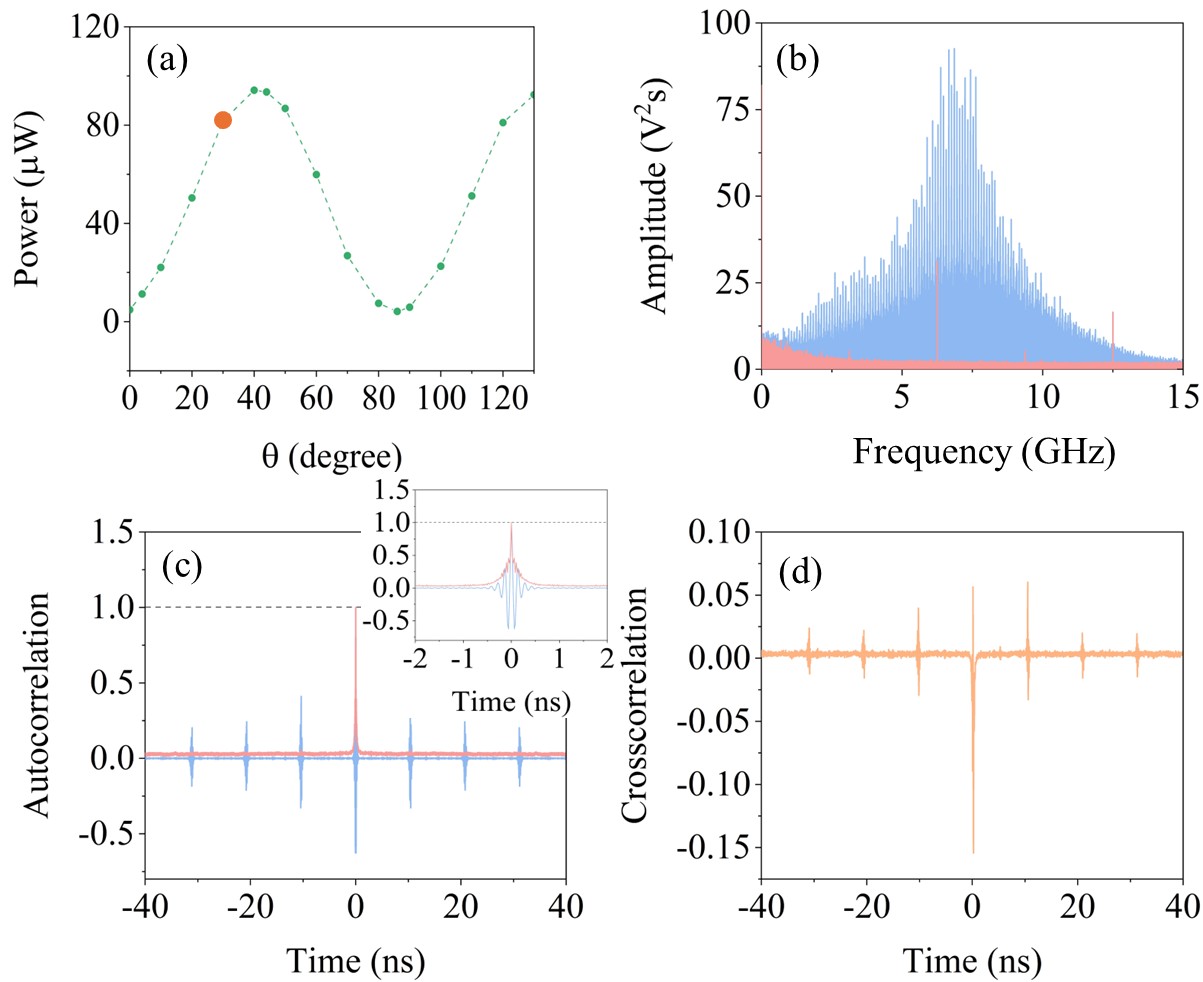}
  \caption{Dynamical signatures of polarization mode competition under optical feedback. (a) Measured optical feedback power as a function of the $\theta$/2-waveplate rotation angle ($\theta$). The orange circle marks the operating point ($\theta = 30^\circ$, $P_{fb} \approx 82 \mu W$) for the subsequent analysis. (b) RF spectra of the TE (blue) and TM (red) modes under orthogonal polarization feedback. (c) Autocorrelation functions of the TE (blue) and TM (red) modes. (d) Cross-correlation function between the TE and TM modes.}
  \label{Injection_autocorrelation}
\end{figure}

Fig.~\ref{Injection_autocorrelation}a characterizes the feedback power as a function of the rotation angle of the $\lambda$/2 angle, $\theta$. It is worth to mention that the angle $\theta$ is refereed to the value indicated on the rotator. The feedback power varies periodically with $\theta$, reaching the maximum at $\theta \approx 42^\circ$, and the minimum at $\theta \approx 86^\circ$. The operating point for subsequent analysis, marked by an orange circle, is set to $\theta = 30^\circ$, corresponding to a feedback power of approximately 82 $\mu$W. This point was chosen for initial detailed analysis as it provides a regime with a clearly observable rate of extreme events, prior to exploring the full parameter dependence.

Under this condition ($J = 3.00$ mA, $\theta = 30^\circ$), the introduction of orthogonal polarization feedback dramatically alters the laser dynamics. The RF spectrum of the TE mode, blue in Fig.~\ref{Injection_autocorrelation}b, undergoes a transition from a single relaxation oscillation peak (c.f. Fig.~\ref{Setup_freerunning}c) to a highly structured comb of sharp, equally spaced peaks (mode spacing: 98.5 MHz, corresponding to a 10.2 ns cavity round-trip) superimposed on a broad background. This structured spectrum unambiguously confirms the dominance of polarization-selective optical feedback. 
Concurrently, the TM mode's RF spectrum (Fig.~\ref{Injection_autocorrelation}b, red) is significantly enhanced at low frequencies, indicating its activation far above the free-running noise floor due to the induced mode competition~\cite{panajotov2012optical, gatare2009mapping}.


The distinct dynamical natures of the two modes are quantified by their autocorrelation functions (Fig.~\ref{Setup_freerunning}c). The TE mode's autocorrelation (blue) shows strong, periodic echoes from the external cavity, with a high-resolution view revealing a sharp central peak and damped relaxation oscillations. Conversely, the TM mode's autocorrelation (red) is predominantly flat with a narrow central spike, characteristic of chaotic dynamics. The cross-correlation function (Fig.~\ref{Setup_freerunning}d) reveals a negative central peak at $\tau$ = 0, indicating anti-correlation and providing direct evidence of intense mode competition. The pronounced asymmetry of the cross-correlation reveals a non-reciprocal energy transfer, underscoring the complex, directional nature of the nonlinear coupling under optical injection~\cite{virte2013deterministic}.

\subsection{Emergence and Statistics of Extreme Events}
\begin{figure}[ht!]
\centering
  \includegraphics[width=8.5cm]{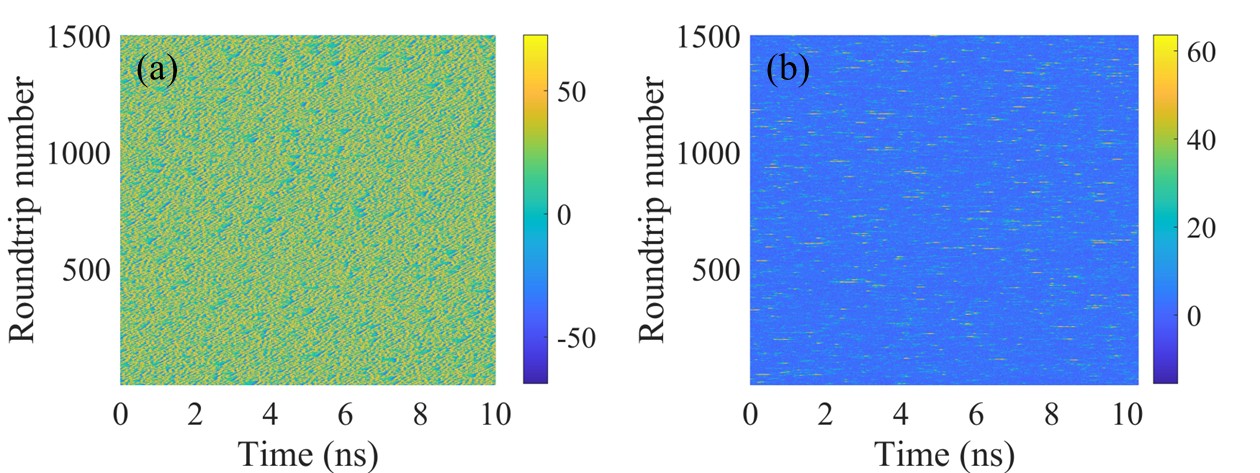}
  \caption{Spatiotemporal dynamics reveal distinct behaviors of TE and TM modes under optical feedback. The horizontal axis represents time over one period ($0-10.2$ ns), and the vertical axis is the roundtrip number. (a) TE mode; (b) TM mode.}
  \label{Spatial_temporal}
\end{figure}

Spatiotemporal diagrams (Fig.~\ref{Spatial_temporal}) contrast the dynamics under feedback. The TE mode (Fig.~\ref{Spatial_temporal}a) exhibits a structured pattern stemming from deterministic cavity interference. The TM mode (Fig.~\ref{Spatial_temporal}b), however, displays stochastic, localized intensity bursts, a signature of extreme events~\cite{pang2024unveiling, akhmediev2016roadmap}. This dichotomy arises from their distinct roles: the TE mode, as the primary lasing mode, is stabilized by coherent feedback, while the TM mode acts as a secondary channel driven into instability via intense gain competition.

\begin{figure}[ht!]
\centering
  \includegraphics[width=8.5cm]{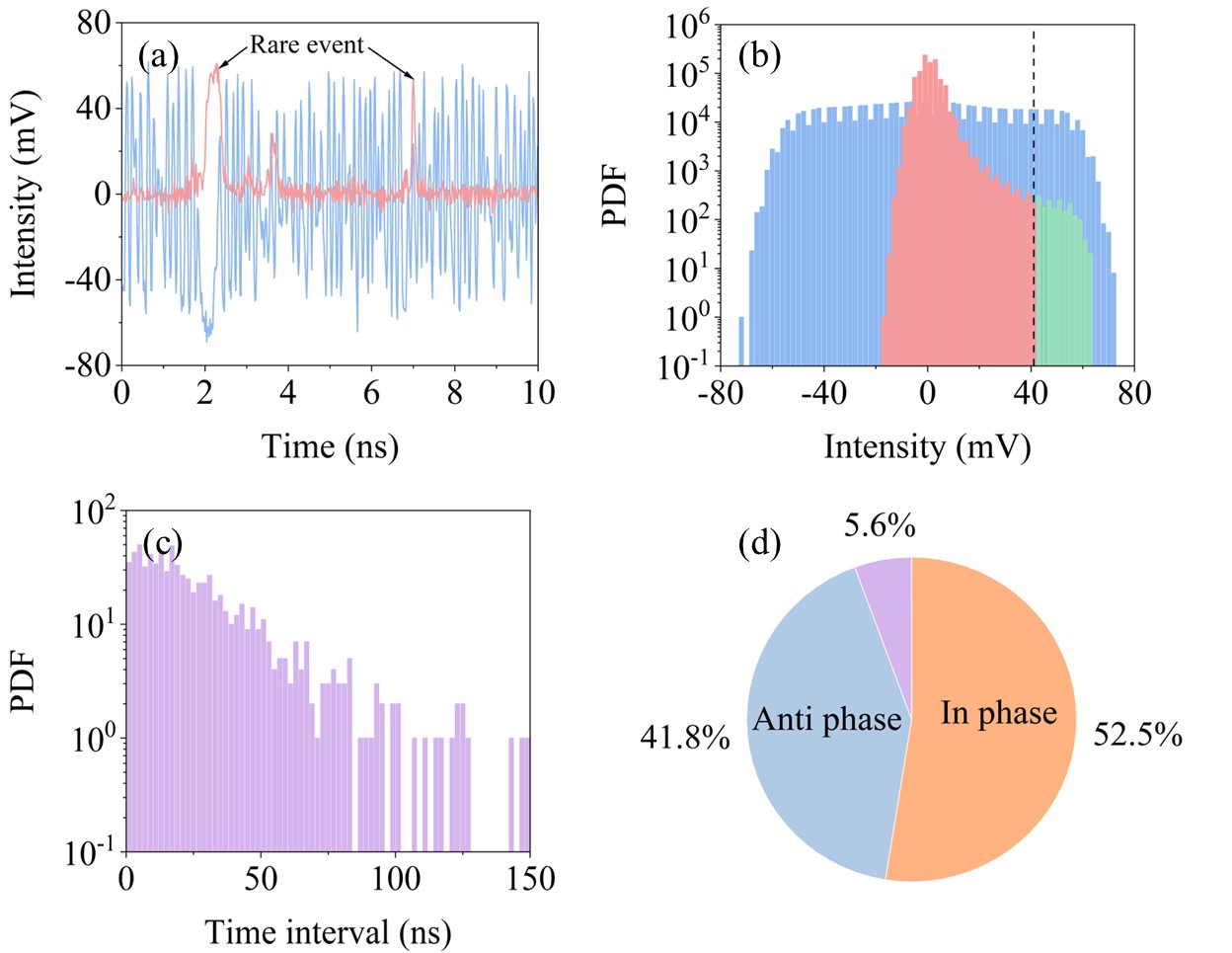}
  \caption{Characterization of extreme events: (a) Time series of the TE (blue) and TM (red) modes, showing high-amplitude chaotic pulses in the TM mode correlated with in-phase or anti-phase counterparts in the TE mode. (b) Log-linear intensity histograms for both modes. The dash line indicates the threshold of the extreme events accordingly to the definition given in the text. Green columns indicate the extreme events. (c) Distribution of time intervals between extreme events, showing broad timing variability. (d) Pie chart quantifying the phase relationship between TE and TM modes during extreme events.}
  \label{Extreme_events}
\end{figure}

A detailed characterization is provided in Fig.~\ref{Extreme_events}. The time series (Fig.~\ref{Extreme_events}a) shows the TE mode undergoing complex oscillations, while the TM mode exhibits sporadic, high-intensity pulses (extreme events) on a low background. Interestingly, each TM extreme event is temporally correlated with a corresponding pulse in the TE channel, manifesting in either anti-phase or in-phase relationships. This indicates a deterministic, nonlinear energy exchange~\cite{heil2001chaos}.

The statistical nature is quantified by the intensity probability distributions of the whole dynamics within 20 $\mu$s (Fig.~\ref{Extreme_events}b). The TE mode's distribution (blue) shows a symmetry structure, consistent with noisy oscillations. The TM mode's distribution (red) deviates dramatically, exhibiting a pronounced heavy tail, which is the definitive statistical signature of extreme events. We define extreme events as those exceeding the threshold $\langle I \rangle + 8\sigma_I$ (where $\langle I \rangle$ and $\sigma_I$ are the average intensity and the corresponding standard deviation).  This high threshold, consistent with other studies in optics~\cite{bonatto2011deterministic}, cleanly isolates the rare, high-amplitude tail from the bulk of the fluctuations. Events populating this tail are highlighted in green~\cite{bonatto2011deterministic}. The calculated kurtosis of the TM intensity distribution is significantly greater than 3 (that of a Gaussian), quantitatively confirming its heavy-tailed nature. The distribution of time intervals between these events (Fig.~\ref{Extreme_events}c) is broad and non-Poissonian, reflecting the complex, chaotic underlying dynamics where the timing of events, while deterministic in origin, appears stochastic. 

Fig.~\ref{Extreme_events}d quantifies the phase relationship at the instant of TM extreme events. The analysis reveals that these events are predominantly partitioned into two distinct classes: 52.5$\%$ of all extreme events occur in-phase with the TE mode, while 41.8$\%$ occur in anti-phase. The remaining 5.6$\%$ exhibit more complex phase relationships. This near-balanced, bipolar distribution is a signature of robust mode competition, indicating that the extreme events are triggered by a nonlinear energy transfer process where energy is rapidly exchanged between the orthogonal modes. The comparable probabilities of in-phase and anti-phase occurrences suggest a system poised near an instability, capable of being perturbed into either state of energy exchange.

\subsection{External Control via Feedback Polarization}
\begin{figure*}[ht!]
\centering
  \includegraphics[width=12.5cm]{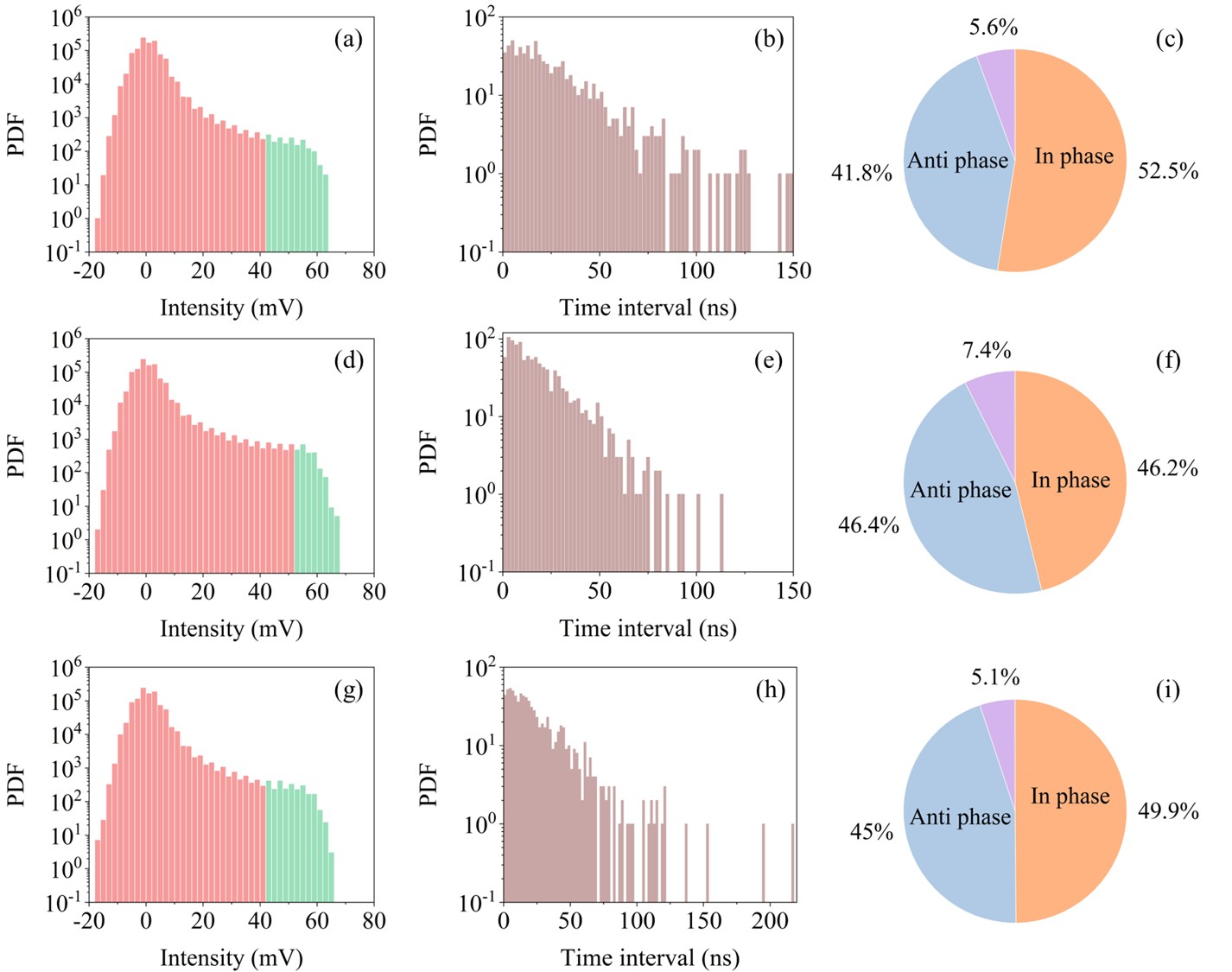}
  \caption{Controlling extreme event statistics via feedback polarization tuning. (a)-(c), Optimal regime at $\theta = 30^\circ$: (a), Intensity probability distribution of the TM mode showing a pronounced heavy tail. Extreme events, defined as intensities exceeding $\langle I\rangle+8\sigma_{I}$, are highlighted in green. (b), Distribution of time intervals between consecutive extreme events, displaying concentrated timing. (c), Phase correlation analysis revealing balanced in-phase and anti-phase relationships with the TE mode. (d)-(f), Detuned regime at $\theta = 40^\circ$: (d), Intensity distribution showing reduced heavy-tail statistics. (e), Broader distribution of inter-event intervals indicating more sporadic timing. (f), Shifted phase correlation suggesting perturbed energy exchange symmetry. (g)-(i), Detuned regime at $\theta = 50^\circ$.}
  \label{Tunable}
\end{figure*}

Fig.~\ref{Tunable} systematically investigates the role of the optical feedback polarization state— controlled by the half-waveplate angle $\theta$ -- as a key parameter for tuning the generation of extreme events. By altering $\theta$, we modulate the polarization orientation and phase of the light reinjected into the VCSEL cavity, thereby modulating the nonlinear coupling strength and the ensuing mode competition between the TE and TM modes. The results reveal a striking non-monotonic dependence of the extreme event dynamics on this parameter.

At $\theta = 30^\circ$, the system operates in a regime of moderate nonlinear coupling, yielding 750 extreme events (within 20 $\mu$s) with a maximum intensity of 63.6 mV (Fig.~\ref{Tunable}a). The distribution of time intervals between these events is relatively broad, extending to approximately 150 ns (Fig.~\ref{Tunable}b), which signifies a substantial stochastic component in their triggering. The phase relationship with the TE mode is nearly balanced, with 46.4$\%$ of events in anti-phase and 46.2$\%$ in-phase (Fig.~\ref{Tunable}c). This balance points to a regime of strong, bidirectional energy exchange where the system does not favor a single energy transfer pathway.

Strikingly, when $\theta$ is tuned to $\theta = 40^\circ$, the feedback conditions become optimal for destabilizing the laser and exciting extreme dynamics. The event count surges to 1086 and the maximum intensity rises to 67.6 mV (Fig.~\ref{Tunable}d), indicating a heightened level of perturbation. Concurrently, the time interval distribution becomes significantly more concentrated, with the maximum interval reduced to $\sim$113 ns (Fig.~\ref{Tunable}e). This shift denotes a higher repetition rate and a transition towards a more deterministic, or ``resonant'', driving process underlying the extreme event generation. The phase correlation remains strongly balanced (46.4$\%$ anti-phase, 46.2$\%$ in-phase, Fig.~\ref{Tunable}f), confirming that the enhanced dynamics are still governed by robust, competitive energy exchange.

Further increasing $\theta$ to $\theta = 50^\circ$ suppresses the extreme dynamics, indicating that the system is moving away from the optimal coupling condition. The event count drops to 803 and the maximum intensity falls to 65.2 mV (Fig.~\ref{Tunable}g). The time interval distribution broadens again, developing a long tail extending to $\sim$220 ns (Fig.~\ref{Tunable}h), which signals a return to a more sporadic and less deterministic generation regime. The phase correlation shows a slight but notable shift (49.9$\%$ in-phase, 45.0$\%$ anti-phase, Fig.~\ref{Tunable}i), suggesting a perturbation in the symmetry of the nonlinear energy transfer, potentially favouring one pathway over the other as the feedback is detuned. 

The observed non-monotonic behaviour--where extreme events are maximized at an intermediate $\theta$ of 40$^\circ$--is a typical signature of a nonlinear resonance. We hypothesize that this optimal angle corresponds to a specific polarization of the reinjected light that maximizes the nonlinear coupling coefficient, effectively driving the polarization mode competition at its most unstable point. This creates the most effective perturbation for triggering high-amplitude pulses through intense, resonant mode competition.

\begin{figure}[ht!]
\centering
  \includegraphics[width=5.5cm]{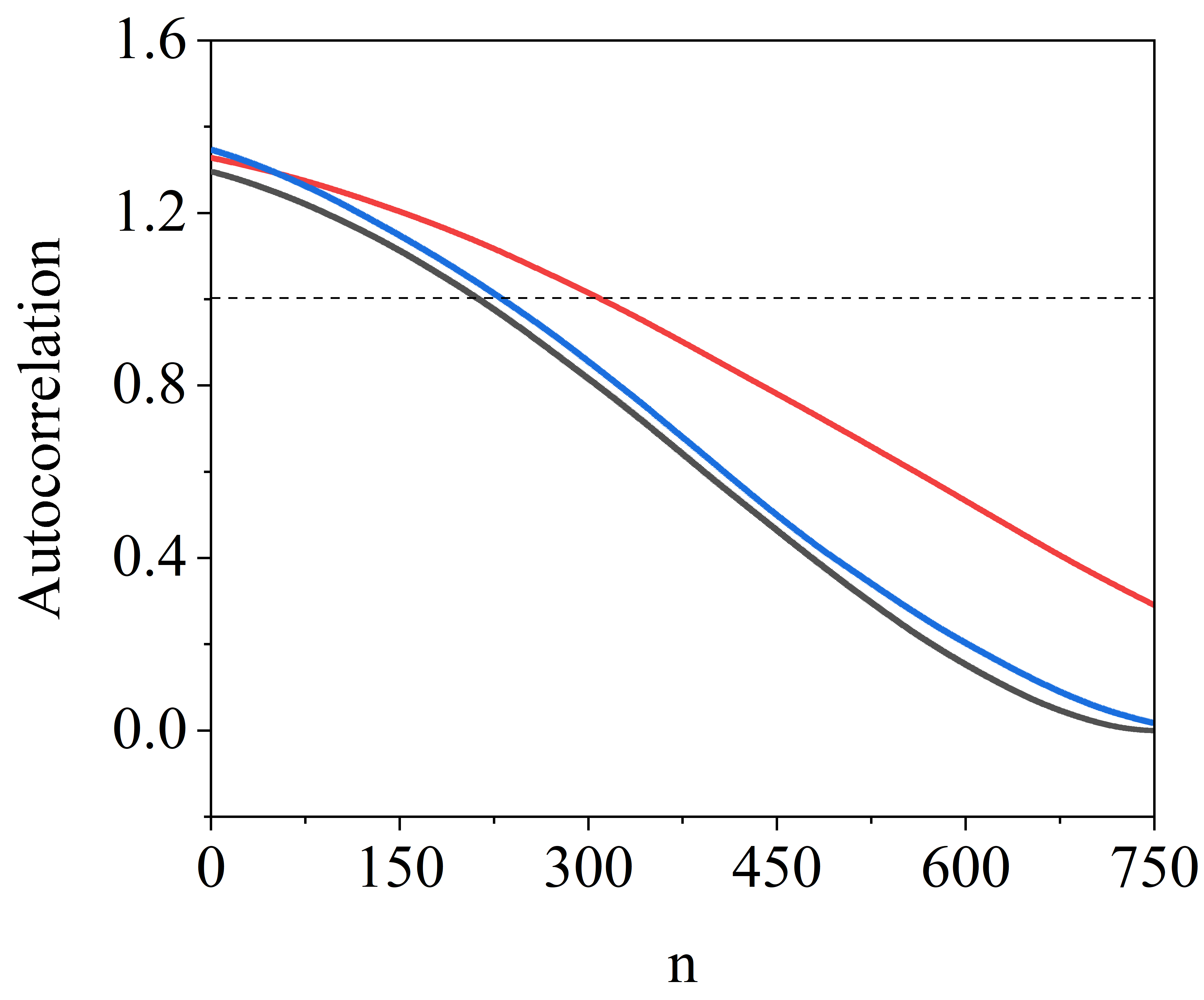}
  \caption{Second-order autocorrelation functions of extreme-event arrival times for $\lambda/2$-plate angles of $\theta = 30^\circ$ (black), $40^\circ$ (red), and $50^\circ$ (blue). The lag is given in units of the number of events.}
  \label{Autocorrelation}
\end{figure}

Fig.~\ref{Autocorrelation} displays the second-order autocorrelation functions of extreme-event arrival times for $\lambda/2$-plate angles of $\theta = 30^\circ$, $40^\circ$, and $50^\circ$. The horizontal axis represents the lag in terms of the number of events, where the autocorrelation at Lag $n$ quantifies the correlation between the arrival time of a given extreme event and that of the $n$-th subsequent event. All curves begin with high positive values (approximately 1.29, 1.33, and 1.34 for $\theta = 30^\circ$, $40^\circ$, and $50^\circ$, respectively), reflecting strong short-range temporal correlations. Notably, the $\theta = 30^\circ$ case exhibits the strongest autocorrelation and slowest decay, indicating the most persistent memory in event timing. In contrast, the $\theta = 50^\circ$ curve decays more rapidly, suggesting diminished correlation persistence, while the $\theta = 40^\circ$ case displays intermediate behavior. This systematic variation demonstrates that the $\lambda/2$-plate angle effectively modulates the memory effects governing extreme-event generation. Furthermore, the smooth, gradual decay of all autocorrelation functions, rather than abrupt drops, suggests that the underlying dynamics involve multiple temporal scales of memory.

These results reveal that the arrival times of extreme events are characterized by long-range temporal correlations and significant memory persistence, diverging markedly from random behavior. The slow decay of autocorrelation over hundreds of events indicates that the timing of each extreme event influences subsequent events far into the sequence. Such structured temporal organization supports the presence of a deterministic and self-organizing process, where extreme events occur in clustered bursts separated by quiescent intervals. The tunability of this clustering behavior via the waveplate angle underscores the role of coherent physical mechanisms in extreme event generation, with higher coherence at $30^\circ$ and a trend toward stochasticity at $50^\circ$. These findings affirm that extreme events in the system are not merely anomalous outliers, but arise from correlated dynamics with a well-defined correlation length in event space, enabling intermediate-term predictability.

\section{Conclusion}
In conclusion, we have introduced and experimentally demonstrated a novel approach for the deterministic generation and control of optical extreme events in a VCSEL using polarization-engineered feedback. By integrating a $\lambda$/2-waveplate within a polarization-selective external cavity, we have achieved precise modulation of the nonlinear interaction between TE and TM modes, leading to the emergence of high-amplitude, heavy-tailed intensity fluctuations in the TM channel, one typical signature of extreme wave dynamics.

We have shown that these extreme events are not random outliers but originate from deterministic energy exchange between polarization modes, as evidenced by their strong bipolar correlation with the TE dynamics and long-range temporal memory. The ability to systematically tune event statistics via the waveplate angle underscores the role of resonant nonlinear coupling in their formation. In particular, the non-monotonic dependence of event rate and intensity on the feedback polarization highlights an optimal regime for extreme event generation, consistent with a nonlinear resonance mechanism.

This work not only advances the fundamental understanding of extreme events in active dissipative systems but also provides a highly configurable experimental platform for future studies—-such as real-time prediction, noise-mediated dynamics, and network synchronization. From an applied perspective, the demonstrated controllability opens pathways for exploiting extreme events in photonic technologies, including optical sensing, random number generation, and signal encryption.

\section*{Acknowledgment}
This work is partially supported by National Natural Science Foundation of China (Grant No. 62475206), Key Research and Development Plan of Shaanxi Province of China (Grant No. 2024GH-ZDXM-42).




\bibliography{biblio}


\end{document}